\documentclass[aps, pra, superscriptaddress, twocolumn, amsfonts, amsmath, amssymb, floatfix]{revtex4-2}
\usepackage{graphicx}
\usepackage{float}
\usepackage{sidecap}
\usepackage{dcolumn}
\usepackage{bm}
\usepackage{epsfig}
\usepackage{braket}
\usepackage{color}
\usepackage{ulem}
\usepackage{amsmath}
\usepackage[colorlinks=true, pdfstartview=FitV, linkcolor=blue, citecolor=blue, urlcolor=blue]{hyperref}
\usepackage[utf8]{inputenc}
\usepackage[T1]{fontenc}

\begin{document}

\title{Quenching dynamics of vortices in spin-orbit coupled Bose-Einstein condensates under a detuning gradient}

\author{Juan Wang}
\affiliation{School of Physics, East China Normal University, Shanghai 200241, China}

\author{Zhenze Fan}
\affiliation{School of Physics, East China Normal University, Shanghai 200241, China}

\author{Yan Li}
\email{yli@phy.ecnu.edu.cn}
\affiliation{School of Physics, East China Normal University, Shanghai 200241, China}

\begin{abstract}
We investigate the ground states and complex dynamical behaviors of vortices in a spin-orbit coupled Bose-Einstein condensate (BEC) subjected to a position-dependent detuning. By scanning the detuning gradient and solving the coupled Gross-Pitaevskii equations, we obtain static vortex lattice structures containing 1 to 6 vortices. When the detuning gradient is quenched below its initial value, these vortex lattices exhibit novel periodic rotation motion, with the stability persisting for up to 1000 ms. Notably, twin vortices demonstrate two distinct modes depending on the gradient: a scissors-like oscillation or a unidirectional clockwise rotation. We quantitatively analyze this dynamics, establishing a direct correspondence between the rotation and the underlying magnetic field gradients experienced by the condensate. In contrast, quenching the detuning gradient beyond its initial value triggers the nucleation of additional vortices. Our findings not only elucidate the rich response of vortices to synthetic magnetic fields but also suggest practical insights for engineering ultra-sensitive, BEC-based magnetic field gradiometers. 
\end{abstract}

\maketitle


\section{Introduction}
Quantum quench describes the nonequilibrium dynamics induced by a sudden change in the Hamiltonian's coupling parameters \cite{PhysRevA.84.023601,RevModPhys.83.863,PhysRevA.99.053609,cplquench,qlx5-dqr3,8cdt-n3l2,w1n4-d3lf}. The existence of quantized vortices serves as conclusive evidence for superfluid properties of Bose-Einstein condensates (BEC) \cite{vortices1999,PhysRevA.67.033610,PhysRevA.67.041602}. The high controllability of BEC provides an ideal platform for studying vortices in quantum and condensed matter physics \cite{science1995,PhysRevLett1995,rotating2000,lattice2001,PhysRevLett.120.186103,science.1177980,PhysRevA.94.053611}. Up to now, a variety of methods have been experimentally realized to generate vortices in BEC, such as phase imprinting \cite{phaseimpring1999,phaseimpring2000,phaseimpring2004}, laser stirring \cite{laser2001,laser2006,laser2016}, rotating external trapping potential \cite{rotating2000,rotating2001,rotating2010}, dragging obstacles \cite{PRL2010,drag2015}, and artificial gauge fields \cite{syn2009,syn20092,PhysRevA.102.011303}. 
Among these, synthetic spin-orbit coupling (SOC) offers a particularly versatile approach to engineer topological states and manipulate vortex dynamics\cite{soc2011,PhysRevA.84.063604}. In recent years, the explorations of quantized vortices have expanded from the fundamental static and dynamic vortex structures to complex collective phenomena, including vortex dipoles \cite{PRL2010,PRX2023}, vortex lattices \cite{sci2024,PhysRevLett.133.143401}, vortex solitons \cite{cpl2025,PhysRevE.111.024205}, vortex droplets \cite{PRA2018,SANJAY2025116441,771t-rkdj}, K\'{a}rm\'{a}n vortex streets \cite{PhysRevLett.104.150404,PhysRevE.102.032217}, and quantum turbulence \cite{SciPost,PhysRevA.106.023306}.

The realization of synthetic spin-orbit coupling (SOC) has further expanded the toolbox for vortex generation and manipulation in ultracold atoms \cite{syn2009,soc2011,PhysRevA.84.063604,PhysRevA.84.063624,PhysRevA.91.043629,WOS:000387551500008,PhysRevE.94.032202,PhysRevLett.122.123201,PhysRevA.106.063321,NP2025,PhysRevA.95.043605,PhysRevA.104.053325,ZHONG2024115590,kb8m-847n}. Unlike conventional methods that rely on rotating traps, SOC enables vortex creation via a position-dependent detuning, which mimics a synthetic magnetic field and eliminates the need for mechanical rotation \cite{syn2009,PhysRevA.84.063604,PhysRevLett.118.145302, PhysRevLett.120.183202}. Such a detuning plays the role of an angular velocity in rotating system \cite{PhysRevLett.118.145302,PhysRevLett.120.183202}. Thus, the creation of vortices can be simply controlled by adjusting the detuning gradient without rotation. Current researches primarily focus on the physical properties and dynamics of vortex-free condensates in synthetic magnetic fields, including intriguing collective excitations and spin-dependent expansion dynamics \cite{PhysRevLett.120.183202,PhysRevA.108.053316,PhysRevA.110.043307,PhysRevResearch.7.013219,pra20251}. 

However, the dynamical behavior of quantized vortices under such synthetic fields, especially their response to quenched detuning gradients, remains largely unexplored. The tunable vortex dynamics in this system, coupled with its sensitivity to detuning gradients, suggest its potential as a novel quantum sensor for magnetic field gradients. These researches on sensing through the measurement of quantum effects are booming. For instance, detecting density fluctuations in a BEC due to the deformation of the trapping potential can be used as a ultracold atomic magnetometer \cite{nature440,apl2006,prappled2017}; measuring the rotation frequency of the minimal atomic density line can serve as a quantum sensor of interactions or scalar magnetic fields \cite{njp2018}; the sensitivity of a ring BEC to the rotation of the optical lattice can be applied as a rotation sensor \cite{PhysRevA.109.023524}.

\begin{figure*}[htbp]
\centering
\includegraphics[width=0.89\textwidth]{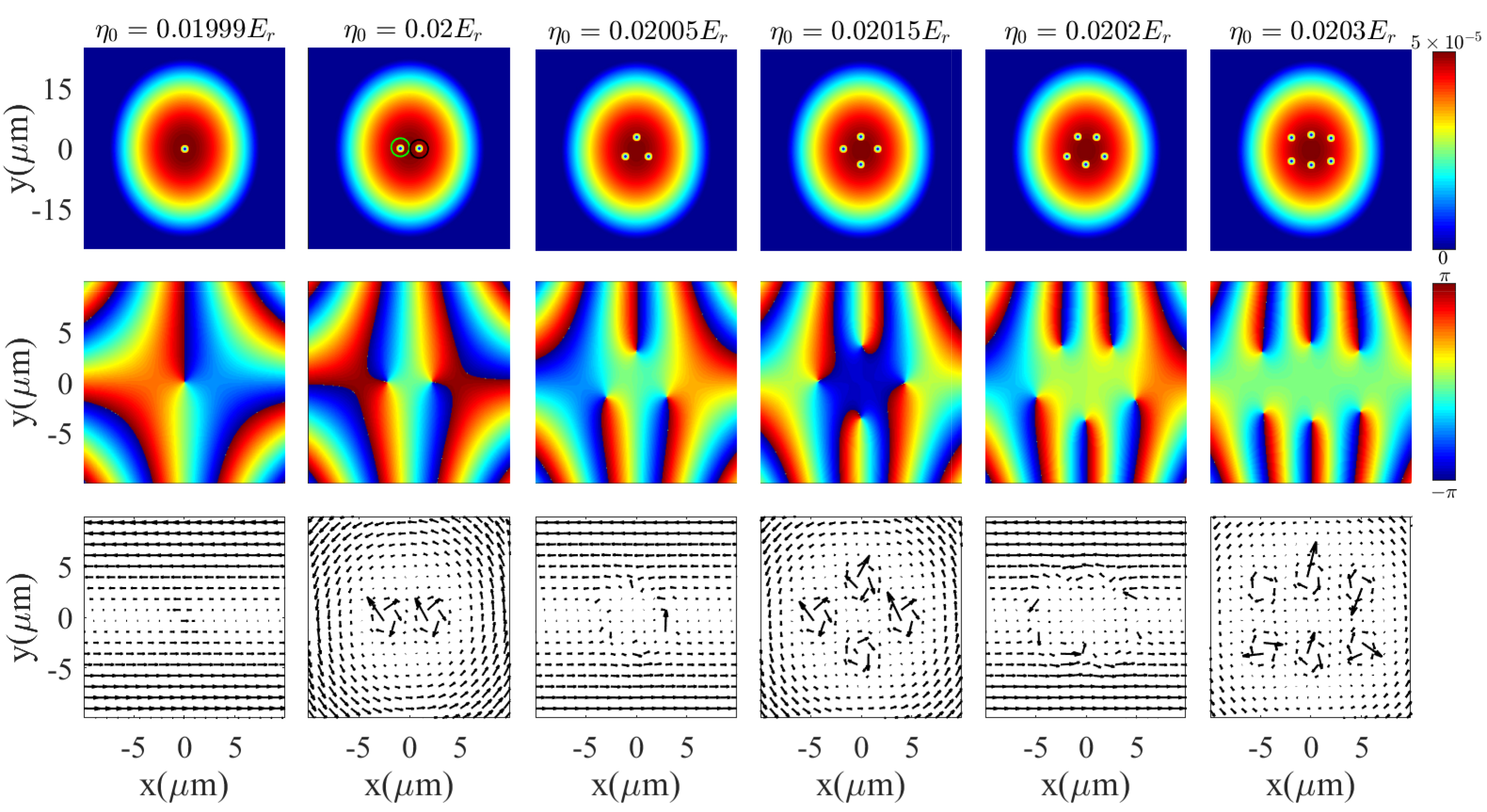}
\caption{The ground states of the vortex lattices. The first row shows the total density distribution of condensates containing 1 to 6 vortices from left to right with $\eta_{0} =0.01999,0.02,0.02005,0.02015,0.0202,0.0203E_{r} $. The second and third rows show the corresponding phase and total velocity field, respectively. The results are obtained from GP simulations with atom number $N=5\times 10^{5} $, Raman coupling $\Omega =10E_{r}$, and isotropic trapping frequencies $(\omega _{x} ,\omega _{y})=2\pi \times (50,50)$ Hz. The green and black circles in two-vortex state mark the vortex position on the $x$ negative semi-axis and positive semi-axis at the initial time, respectively.}
\label{fig:groundstate}
\end{figure*}

In this work, we investigate the influence of position-dependent detuning on the vortex dynamics by numerically solving the coupled Gross-Pitaevskii (GP) equations. When quenching detuning gradient below a certain value, we find that the vortex lattices exhibit periodic rotational dynamics with remarkable stability persisting up to 1000 ms. Notably, the twin vortices exhibit distinct oscillation angles and rotation periods depending on the quenched detuning gradient, reflecting the response to the gradient magnetic field. By fitting the numerical results, we quantify the relation between the dynamics of twin vortices and the magnetic field gradient, which provides a promising foundation for future experimental research and technological development in magnetic field gradient measurements. Moreover, when quenching the detuning gradient above its initial value, additional vortices appear due to the sudden enhancement of the synthetic magnetic field.

This paper is structured as follows. In Sec.~\ref{sec-model}, we introduce the theoretical model and analyze the ground states of vortex structures by solving coupled GP equations. In Sec.~\ref{sec-dynamics}, we study the quenching dynamics of various vortex configurations. Section~\ref{sec-conclusion} is our conclusion.

\section{model} 
\label{sec-model}
We consider the following single-particle Hamiltonian for a SOC BEC subject to a synthetic magnetic field \cite{syn2009,PhysRevLett.118.145302,PhysRevLett.120.183202}:
\begin{align}
\hat{H}_{sp}  = \frac{1}{2m}( \hat{p} -\hbar k_{r} \sigma _{z}\hat{e}_{x}  ) ^{2}+  V_{\mathrm{trap} } -\frac{\Omega }{2} \sigma _{x} - \delta(y)\sigma _{z},  
\label{eq:Hamiltonian}
\end{align}
where $ \hat{p}=-i\hbar \nabla$ is the canonical momentum, $\hbar k_{r} $ is the recoil momentum generated by the Raman laser. $V_{\mathrm{trap} } =\frac{m}{2} (\omega _{x}^{2}x^{2} + \omega _{y}^{2}y^{2})$ is the two-dimensional (2D) harmonic potential with trapping frequencies $\omega _{x,y}$, where $m$ is the atomic mass. In this work, we focus on the zero-momentum phase characterized by $\Omega > \Omega_{c} \equiv 4E_{r}  $, where $\Omega $ is the Raman coupling strength, $E_{r} =\hbar ^{2} k_{r}^{2} /2m$ represents the recoil energy.  The last term $\delta (y)=\eta k_{r} y$ corresponds to a position-dependent detuning, which can be generated by applying a spatially dependent magnetic field in experiment. Interestingly, with the detuning gradient $\eta$ increasing to a critical value, quantized vortices begin to appear.

The nonlinear interaction between atoms can be taken into account through the mean-field approximation, and the Hamiltonian is given by:
\begin{equation}
\hat{H} _{int} =\begin{pmatrix}g_{11}|\psi _{1} |^{2} + g_{12}|\psi _{2} |^{2}
  & 0 \\
 0 & g_{12}|\psi _{1} |^{2}+ g_{22}|\psi _{2} |^{2}
\end{pmatrix}.  
\label{eq:interaction}
\end{equation}
$\psi _{j=1,2} $ denotes the order parameter of each component, satisfying the normalization condition $\int dxdy (|\psi _{1} |^{2} +|\psi _{2} |^{2})=N$, where $N$ is the total number of atoms. The effective 2D interaction constants $g_{ij} =\sqrt{8\pi } (\hbar ^{2} /m)a_{ij} /a_{z} $ are determined from the s-wave scattering lengths $a_{ij}$ in different spin channels and the transverse harmonic length $a_{z} =\sqrt{\hbar /m\omega _{z} } $. For simplicity, we consider isotropic coupling constants $g_{ij} =g$. In this regime, the coupled continuity equations for the components can be expressed as:
\begin{subequations}
\begin{align}
\frac{\partial |\psi _{1}|^{2}  }{\partial t} = &\nabla \cdot [\frac{\hbar }{2mi} (\psi _{1}\nabla\psi _{1}^{\ast }-\psi _{1}^{\ast } \nabla\psi _{1})  ] +\frac{\hbar k_{r} }{m}\nabla_{x}|\psi _{1} |^{2} \notag\\
& + \frac{\Omega }{2i\hbar }(  \psi _{1}\psi _{2}^{\ast }-\psi _{1}^{\ast }\psi _{2}), \label{Za}
\\
\frac{\partial |\psi _{2}|^{2}  }{\partial t} =&  \nabla \cdot [\frac{\hbar }{2mi} (\psi _{2}\nabla\psi _{2}^{\ast }-\psi _{2}^{\ast } \nabla\psi _{2} ) ]-\frac{\hbar k_{r} }{m}\nabla_{x}|\psi _{2} |^{2} \notag\\
 &+\frac{\Omega }{2i\hbar }(  \psi _{2}\psi _{1}^{\ast }-\psi _{2}^{\ast }\psi _{1}).\label{Zb}
\end{align}
\end{subequations}
Equations~(\ref{Za}) and (\ref{Zb}) have the form of continuity equations for particle density, $n_{1} =|\psi _{1} |^{2} $, $n_{2} =|\psi _{2} |^{2} $, total density $n=n_{1} +n_{2}$, spin density $s_{z}=n_{1} -n_{2}$. Then we get the simplified equations:
\begin{align}
\label{eq:continuity22}
\frac{\partial n}{\partial t} +\nabla \cdot (n\mathbf{v}  )& = 0,
\end{align}
\begin{align}
\label{eq:vc}
\mathbf{v}=\mathbf{v}^{c} +\mathbf{v}^{s}.
\end{align}
 The total velocity field $\mathbf{v}$ includes the canonical velocity field $\mathbf{v}^{c} =\frac{\hbar }{2mi} \frac{(\psi _{1}^{\ast }\nabla\psi _{1} - \psi _{1}\nabla\psi _{1}^{\ast }+\psi _{2}^{\ast }\nabla\psi _{2} - \psi _{2}\nabla\psi _{2}^{\ast } )}{n}$ and a novel term associated with the spin density $\mathbf{v} ^{s} =-\frac{\hbar k_{r}}{m}\frac{s_{z}}{n}\hat{e} _{x}$.

\begin{figure}[t!]
\includegraphics[width=1\linewidth]{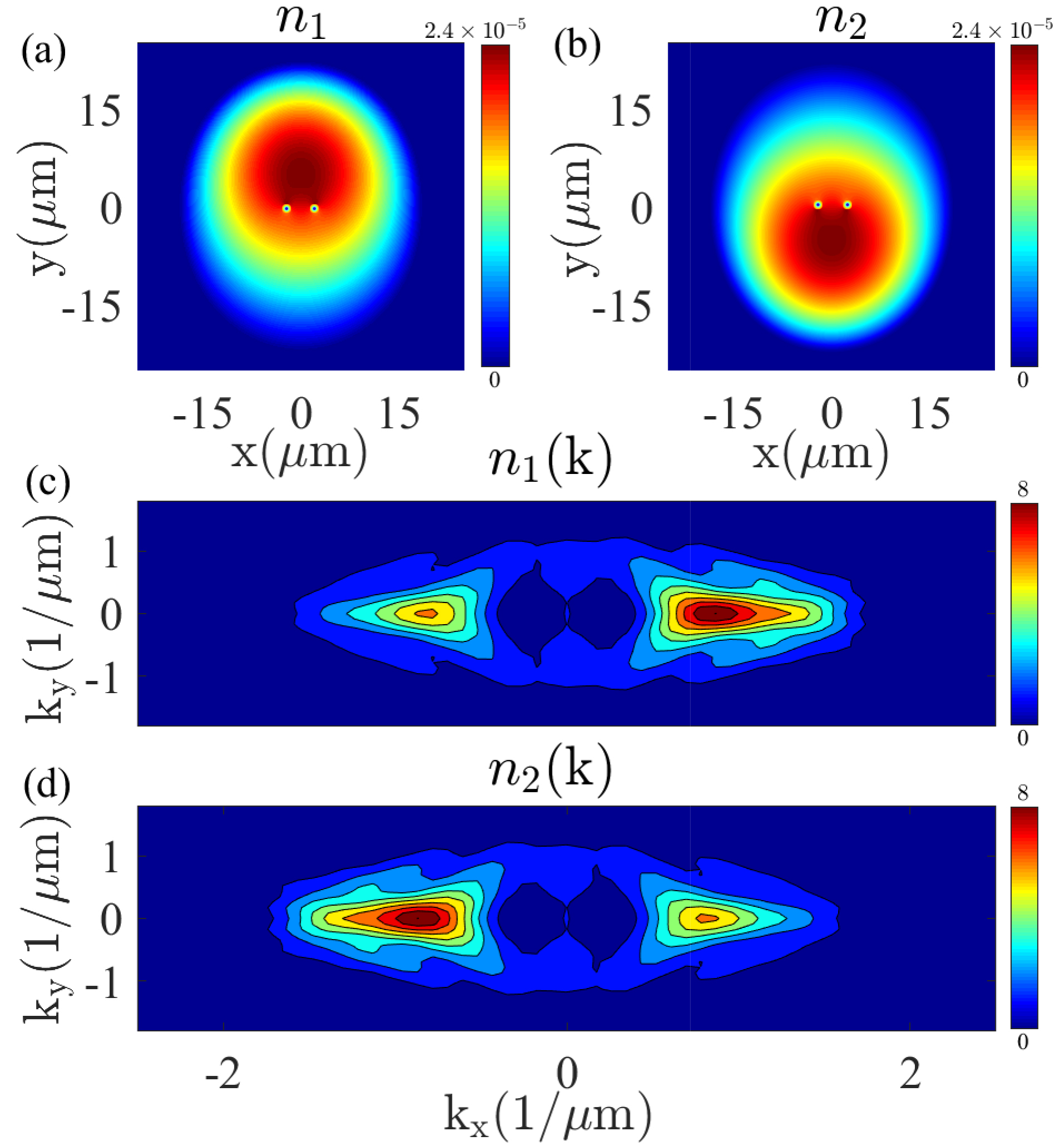} 
\centering
\caption{The ground state of each component containing two vortices. (a) and (b) show the density profile of component $n_{1}$ and $n_{2}$, respectively. The corresponding momentum distribution is shown in (c) and (d), where $n_{1} (k)=\mathcal{F}(n_{1})$, $n_{2} (k)=\mathcal{F}(n_{2})$. The system parameters are the same as Fig.~\ref {fig:groundstate}. }
\label{fig:nAnB}
\end{figure}

 \begin{figure}[t!]
\includegraphics[width=1\linewidth]{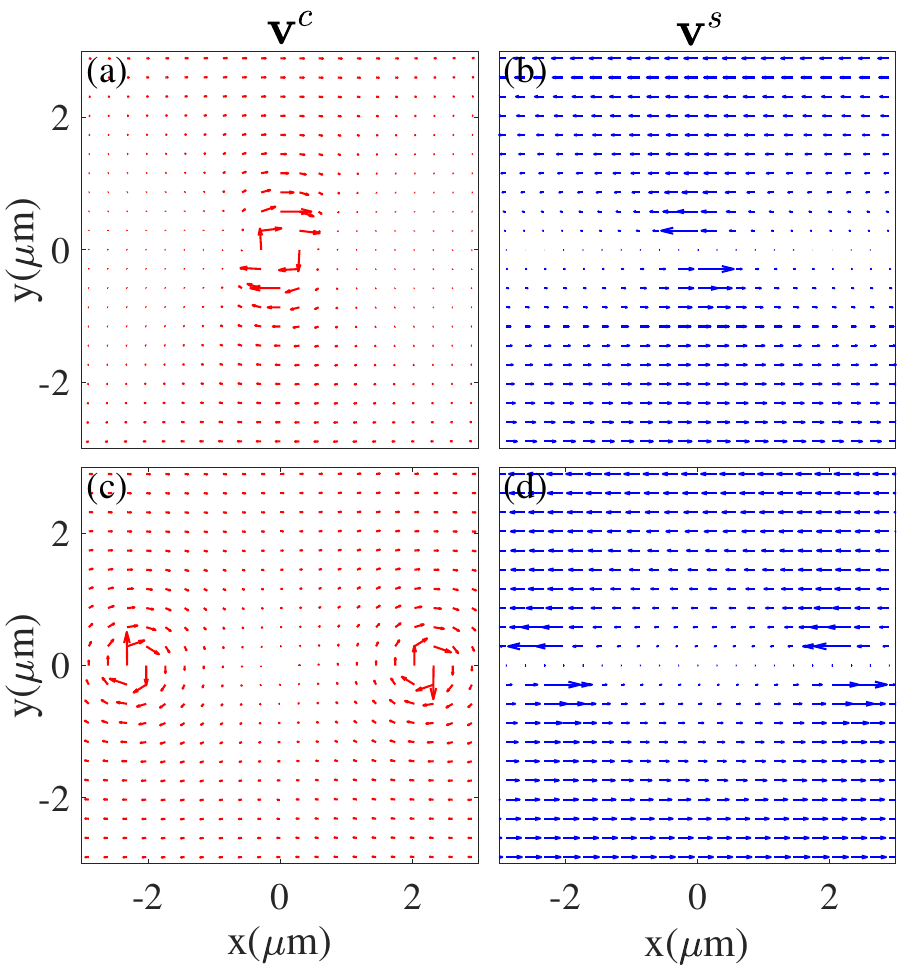} 
\centering
\caption{The velocity field distribution near the vortex core. (a) and (b) show the canonical contribution and the spin contribution to the total velocity field of single-vortex state with $\eta_{0} =0.01999E_{r} $, respectively. (c) and (d) correspond to two-vortex state with $\eta_{0} =0.02E_{r} $. The size of the arrows reflects the magnitude of the velocity field, where long arrows represent large velocities and short ones represent small velocities. }
\label{fig:velocity}
\end{figure}

The detuning gradient controls the strength of the effective magnetic field. At equilibrium, through imaginary time evolution of the coupled GP equations, we scan the detuning gradient for the same atom number and Raman coupling. Vortices emerge at a critical detuning gradient $0.01999 E_{r}$. Figure~\ref{fig:groundstate} shows the vortices' ground states under different initial detuning gradients, including 1 to 6 vortices with lattice structures. The corresponding phase around each vortex core changes $-2\pi $ (with the topological charge $S=-1$), and the density at the center of the vortex is zero. For the zero-momentum phase, one might expect the momentum distribution to be concentrated around a single point at $k=0$. However, this is not necessarily true in the presence of a detuning gradient ($\eta \ne 0$). For example, in the case of two-vortex state, the density profile and momentum distribution of each component are shown in Fig.~\ref {fig:nAnB}. The momentum distributions [Fig.~\ref {fig:nAnB}(c) and (d)] of both components show two peaks, indicating the mixture of two dressed states with opposite quasi-momenta. The dressed state at $k_{x}  >0$ in Fig.~\ref {fig:nAnB}(c) corresponds to a plane-wave phase located at $y>0$ of the real space in Fig.~\ref {fig:nAnB}(a), while the dressed state at $k_{x}  <0$ corresponds to a plane-wave phase located at $y<0$ of the real space. Due to phase mismatch, vortices appear at the overlapping region (i.e., at $y=0$) of the two dressed states. The direction of vorticity is dictated by the two distinct dressed states at $y>0$ and $y<0$. It's very different from that created by a rotating harmonic trap. 

Furthermore, we investigate the velocity field near the vortex core. An interesting phenomenon is that the total velocity field $\mathbf{v}$ of odd numbers of vortices does not really exhibit a visually clear vortex circulation like even numbers of vortices, as shown in Fig.~\ref{fig:groundstate}. Nevertheless, the circulation corresponding to each vortex must be finite. To clearly understand the velocity field of vortex, the canonical contribution $\mathbf{v} ^{c}$ and spin contribution $\mathbf{v} ^{s}$ of the total velocity field for single-and two-vortex configurations are shown in Fig.~\ref{fig:velocity}. Indeed, if we only look at $\mathbf{v} ^{c}$, there are finite circulation and $-2\pi $ phase winding for both one vortex and two vortices [see Figs.~\ref{fig:velocity}(a) and \ref{fig:velocity}(c)], indicating the presence of quantized vortices. In contrast, the spin contribution $\mathbf{v} ^{s}$ only has values in the $x$ component due to the SOC only generating along $x$ direction [see Figs.~\ref{fig:velocity}(b) and \ref{fig:velocity}(d)]. In Fig.~\ref{fig:groundstate}, the finite circulation of the total velocity field for two vortices arises from the superposition of $\mathbf{v} ^{c}$ and $\mathbf{v} ^{s}$. However, the single-vortex configuration exhibits a velocity distribution resembling the spin contribution $\mathbf{v} ^{s}$ in its total velocity field due to the relatively small canonical contribution $\mathbf{v} ^{c}$.

 \begin{figure}[t!]
\includegraphics[width=1\linewidth]{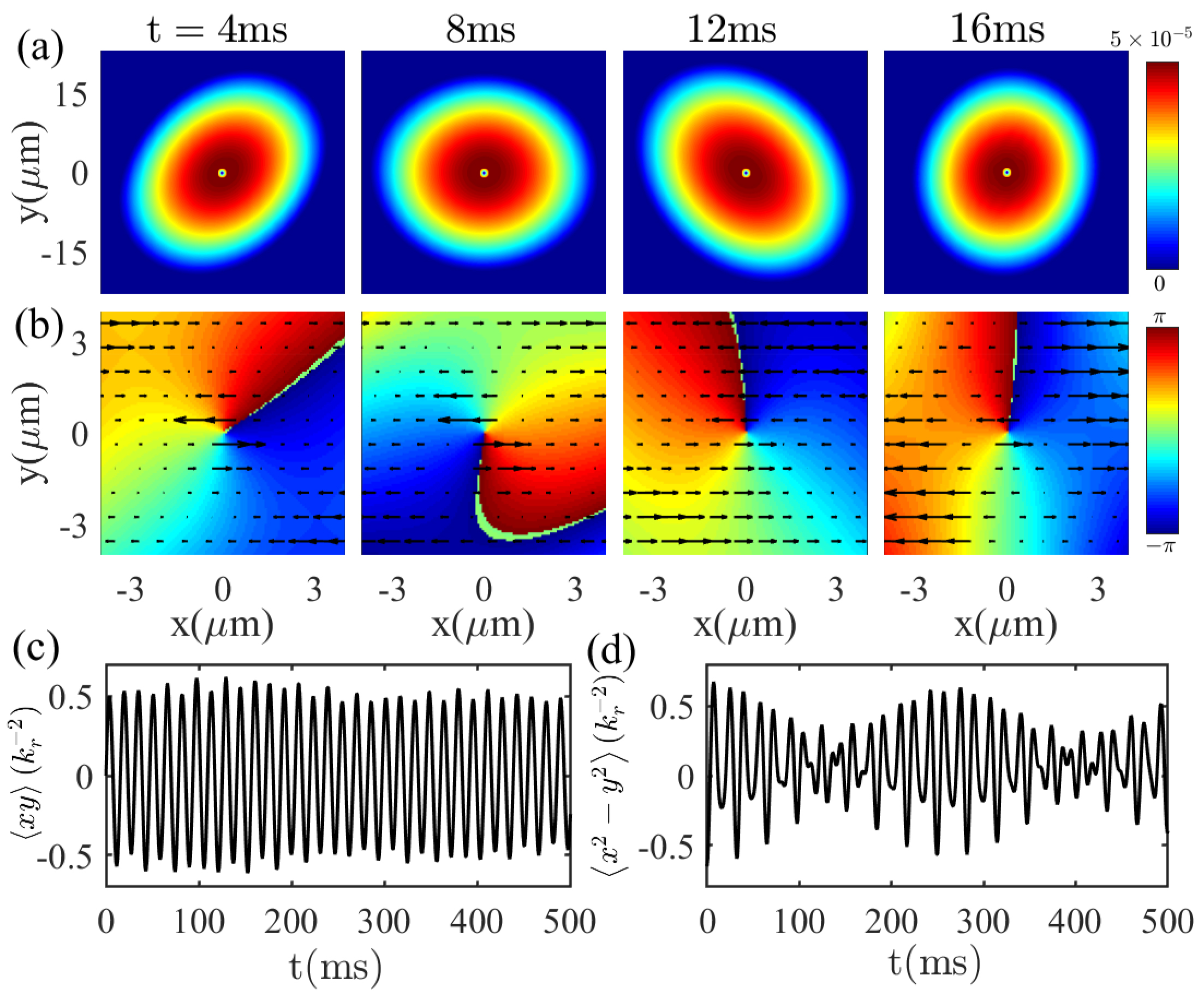} 
\centering
\caption{The dynamical evolution of a single vortex with quenching $\eta =0E_{r}$. (a) shows the total density distribution at different quench times, with the corresponding velocity field (arrows) near the center of vortex core shown in (b). The color map of the background in (b) illustrates the phase distributions of order parameters. (c) and (d) show time evolution of the scissors mode $\left \langle xy \right \rangle $ and quadrupole mode $\left \langle x^{2}-y^{2}   \right \rangle $ obtained from GP equations numerical simulations.}
\label{fig:vortex1}
\end{figure}

\begin{figure}[htbp]
\centering
\includegraphics[width=1\linewidth]{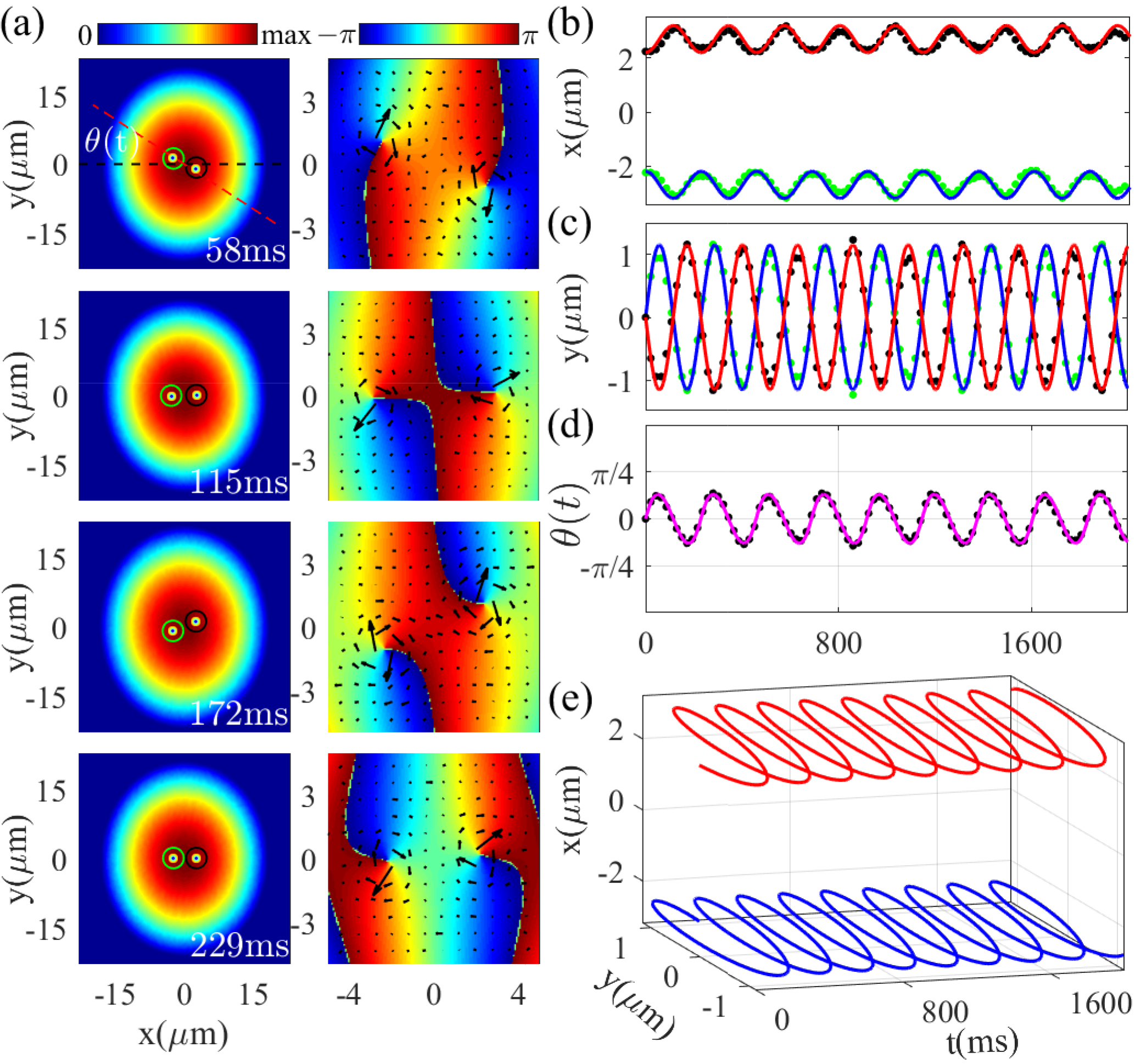}
\caption{The dynamics of two-vortex state with quenching $\eta=0.015E_{r}$. (a) shows the total density profile (left column) and the corresponding phase and velocity field (right column). The green and black circles correspond to the initial vortex positions marked in Fig.~\ref{fig:groundstate}. The rotation angle of two vortices $\theta(t)$ is defined as the deviation from initial position, with measurements restricted to the second and third quadrants of the Cartesian coordinate system. The rotation is positive in the second quadrant and negative in third quadrant, ranging from $(-\frac{\pi }{2},\frac{\pi }{2} ) $. At $t=0$, $\theta(t)=0$. (b) and (c) display the evolution of vortex positions along $x$- and $y$-directions, respectively, where the green symbols (blue solid lines) and black symbols (red solid lines) correspond to the vortex marked by the green and black circles in (a). (d) and (e) show the rotation angle and motion trajectories. In all panels, the solid lines are the fitted results, while the symbols represent results from GP simulations.}
\label{fig:2vortex1}
\end{figure} 

\begin{figure}[htbp]
\centering
\includegraphics[width=1\linewidth]{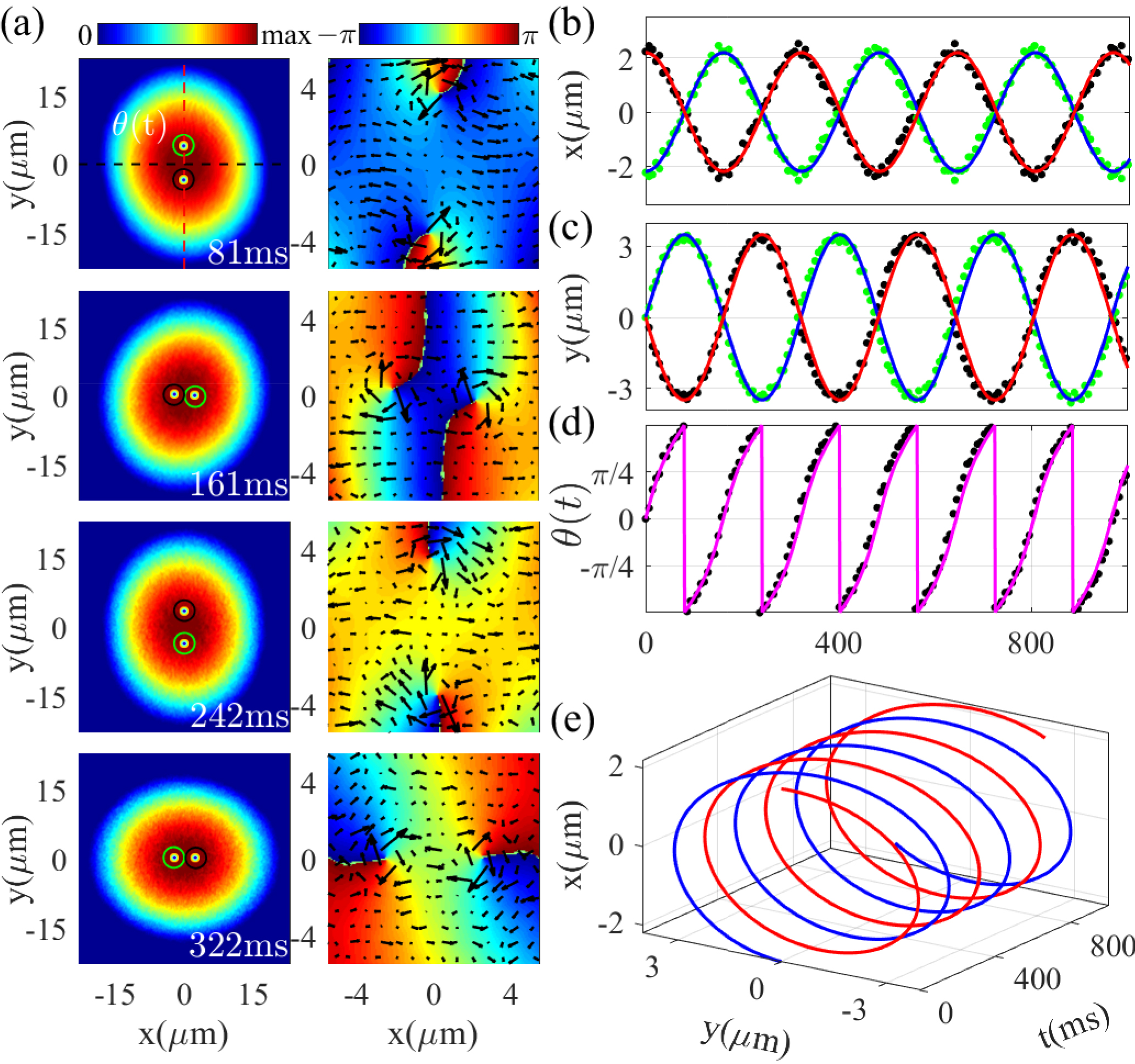}
\caption{The dynamics of two-vortex state with quenching $\eta=0.008E_{r}$. (a) displays the total density (left column) and the corresponding phase and velocity field (right column). (b) and (c) show the evolution of vortices along $x$- and $y$-directions, respectively. The green symbols (blue solid lines) and black symbols (red solid lines) correspond to the evolution of vortices marked by the green and black circles in (a). (d) and (e) show the rotation angle and motion trajectories. The solid lines are the fitted results and the symbols are numerical simulations.}
\label{fig:2vortex2}
\end{figure} 

\section{Dynamics of vortices by quenching detuning gradient}
\label{sec-dynamics}
For different vortex configurations, we investigate their dynamics by numerically solving the coupled GP equations when quenching the detuning gradient to $\eta < \eta _{0} $. Figure~\ref{fig:vortex1} shows the dynamical evolution of a single vortex state with initial detuning gradient $\eta _{0} =0.01999E_{r}$. At equilibrium, the vortex is localized at the center of the condensate $(0,0)\mu $m. When the detuning gradient abruptly switches to $\eta =0E_{r}$, we find that the vortex remains at the center of the condensate $(0,0)\mu $m throughout the time evolution. However, the synthetic magnetic field induces the coupling between the scissors mode and the quadrupole mode \cite{PhysRevA.108.053316,PhysRevResearch.7.013219}, leading to angle oscillations and shape oscillations of the condensate, characterized as $\left \langle xy \right \rangle $ and $\left \langle x^{2}-y^{2}   \right \rangle $, respectively.

For the two-vortex state in initial detuning gradient $\eta _{0} =0.02E_{r}$, the vortices are localized at $(-2.2,0)\mu $m and $(2.2,0)\mu $m. Upon quenching the detuning gradient to $0.015E_{r}$, vortices begin to rotate caused by the synthetic magnetic field. The total density of two-vortex state, the corresponding velocity field and phase are shown in Fig.~\ref{fig:2vortex1}(a). At $t=58$ms, the two-vortex positions are $(-2.7,1.2)\mu $m and $(2.7,-1.2)\mu $m; at $t=115$ms, they are located at $(-3.2,0)\mu $m and $(3.2,0)\mu $m; at $t=172$ms, the vortices move to $(-2.7,-1.2)\mu $m and $(2.7,1.2)\mu $m; and at $t=229$ms, they return to the initial position $(-2.2,0)\mu $m and $(2.2,0)\mu $m. As the quench time increases, the two vortices undergo periodic oscillation analogous to scissors mode [Fig.~\ref{fig:2vortex1}(b) and (c)], with rotational period $229$ms. The rotation angle of two vortices is characterized by $\theta(t)$ [as illustrated in Fig.~\ref{fig:2vortex1}(a)], reaching the maximum value $\theta_{max} =\frac{7\pi }{50} $. To better analyze the motion characteristics of vortices, we fit the numerical results obtained from GP equations and get the motion equations of the two vortices along $x$ and $y$ directions $x_{L,R} (t)=\mp 2.7\pm \frac{1}{2} \cos \frac{2\pi }{229}t $, $y_{L,R} (t)=\pm 1.2\sin \frac{2\pi }{229}t $. Figure~\ref{fig:2vortex1}(e) illustrates the evolution of two-vortex trajectories with time. Our numerical simulations align very well with fitted results in $2000$ms. Indeed, the vortices exhibit a lifetime significantly exceeding $2000$ms, although their trajectories gradually deviate from perfect periodicity beyond this duration.

\begin{figure}[htbp]
\centering
\includegraphics[width=1\linewidth]{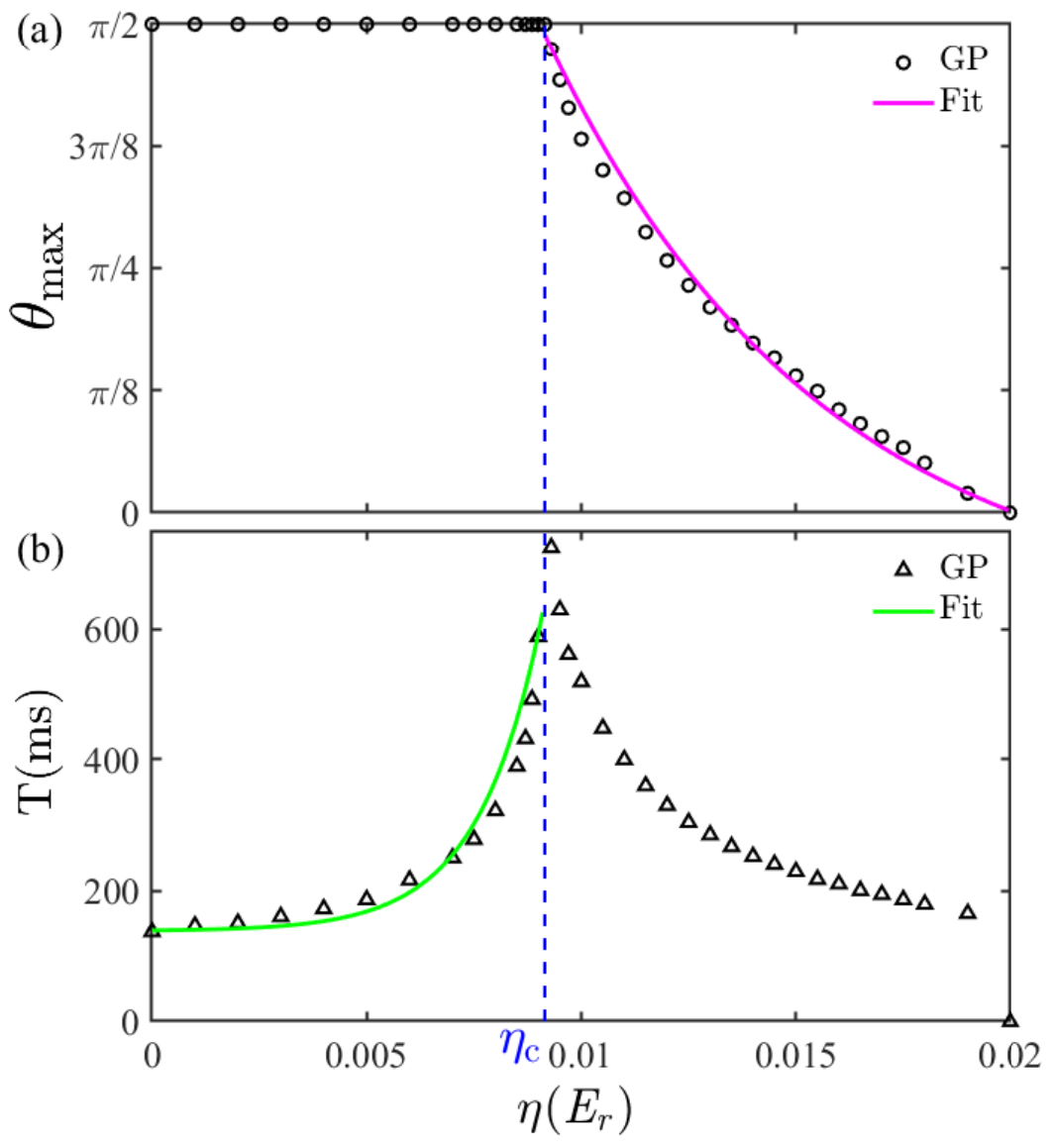}
\caption{(a) Maximum rotation angle $\theta_{max}$ and (b) rotation period $T $ of twin vortices as functions of the quenched detuning gradient $\eta$. The symbols are obtained by numerically solving GP equations, while the solid lines are the fitting results. The blue dashed line corresponds to the quenched critical value $\eta_{c}$.}
\label{fig:theta}
\end{figure} 

Figure~\ref{fig:2vortex2} demonstrates the evolution of the two-vortex state following the quench $\eta =0.008E_{r}$. At quench time $81$ms, the rotation angle of two vortices reaches $\pi/2$, located at $(0,3.5)\mu $m (green) and $(0,-3.5)\mu $m (black). Subsequently, the vortices continue to rotate clockwise, resulting in an abrupt change of $\pi$ in the rotation angle $\theta(t)$. At $161$ms, the rotation angle returns to zero, but the two-vortex positions are interchanged compared to the initial state, being located at $(2.2,0)\mu $m (green) and $(-2.2,0)\mu $m (black). At $242$ms, the rotation angle again reaches $\pi/2$ with the vortex positions at $(0,-3.5)\mu $m (green) and $(0,3.5)\mu $m (black), followed by another abrupt shift to $-\pi/2$. The two-vortex trajectories undergo a full periodic rotation and return to their initial positions at $322$ms, with zero rotation angle. Through fitting the numerical simulations, we obtain the motion equations of two vortices when quenching detuning gradient $\eta =0.008E_{r}$: $x_{L,R} (t)=\mp  2.2\cos \frac{2\pi }{322}t$, $y_{L,R} (t)=\pm 3.5\sin \frac{2\pi }{322}t$. The dynamical stability of two-vortex periodic rotation is maintained for up to $1000$ms. Whether it is scissors-like oscillation or periodic motion, these two vortices with the same topological charge are always symmetrically distributed. Thus, they can be called twin vortices.

Through systematic scanning of detuning gradients, we plot the dependence of both the maximum rotation angle $\theta_{max}$ and the rotational period  $T $ on the quenched detuning gradient, as shown in Fig.~\ref{fig:theta}. For the quenched $\eta$ is close to and less than the initial $\eta_{0}$, the twin vortices exhibit scissors-like rotational oscillation, with both the maximum rotation angle and oscillation period increasing as the quenched detuning gradient decreases. For the quenched $\eta<0.00915E_{r}$, the maximum rotation angle reaches $\theta_{max} = \pi/2$, accompanied by spatial inversion of the two-vortex positions and complete periodic rotation. Furthermore, as the quenched detuning gradient continues to decrease, the rotational period shortens, indicating an increase in vortices' rotational frequency. Taking the quenched detuning gradient $\eta_{c}=0.00915E_{r}$ as the dynamical critical point for twin vortices state, we fit the numerical results: 
\begin{align}
\theta _{max}=0.34e^{-156(\eta -0.02)} -0.34, (0.00915<\eta \le 0.02)   
\label{eq:theta}
\end{align}
\begin{align}
T =e^{680\eta } +136. (0\le \eta <0.00915)   
\label{eq:period}
\end{align}

\begin{figure}[t!]
\centering
\includegraphics[width=1\linewidth]{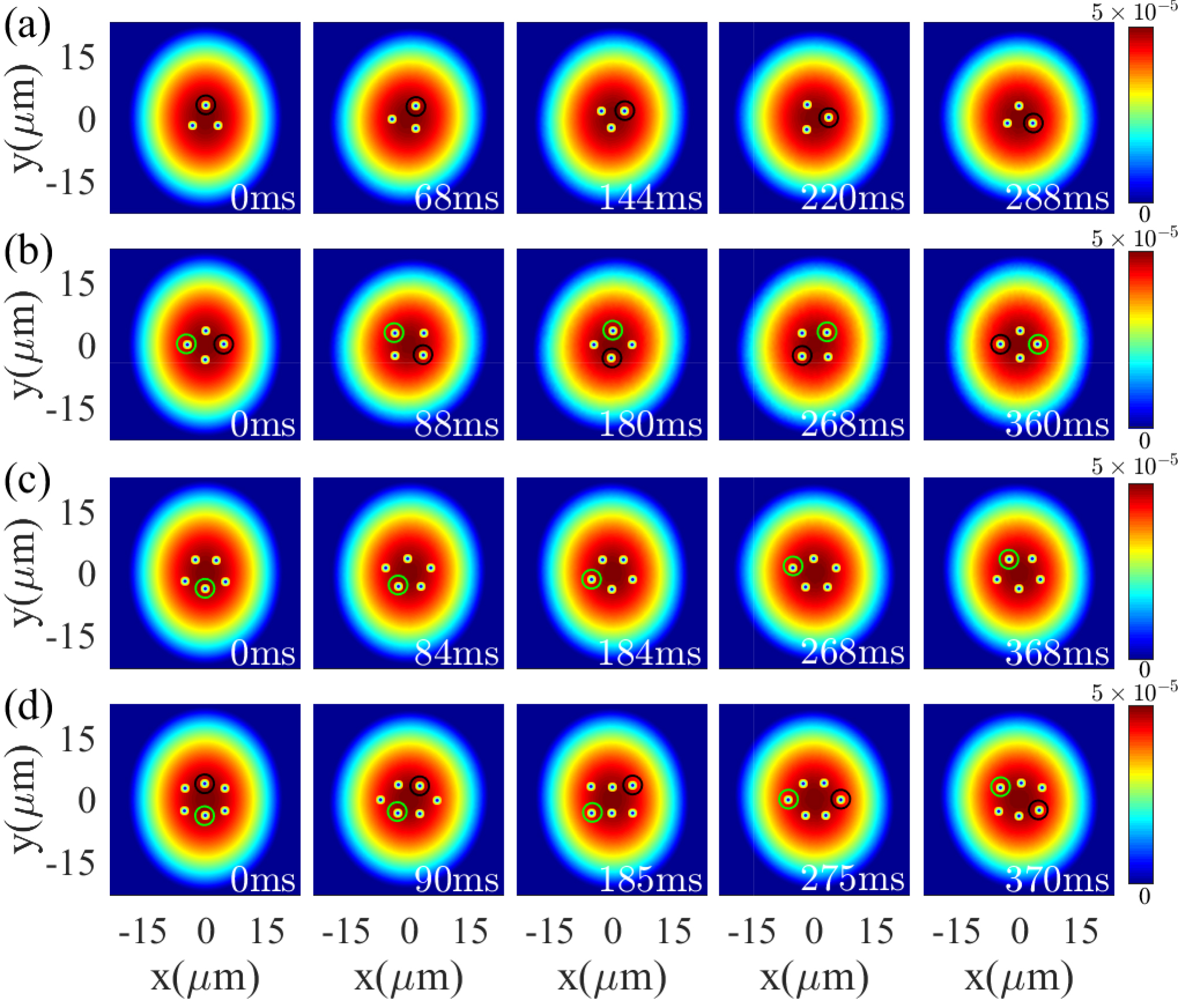}
\caption{Evolution of total density of triangular (a), rhombic (b), pentagonal (c), and hexagonal vortex lattices (d) at different quench times for quenching detuning gradients $\eta= 0.017, 0.016, 0.0175, 0.018E_{r}  $, respectively. The trajectories of vortices marked by green or black circles are displayed in Fig.~\ref{fig:motion}.}
\label{fig:vortex3456}
\end{figure}

\begin{figure}[htbp]
\centering
\includegraphics[width=1\linewidth]{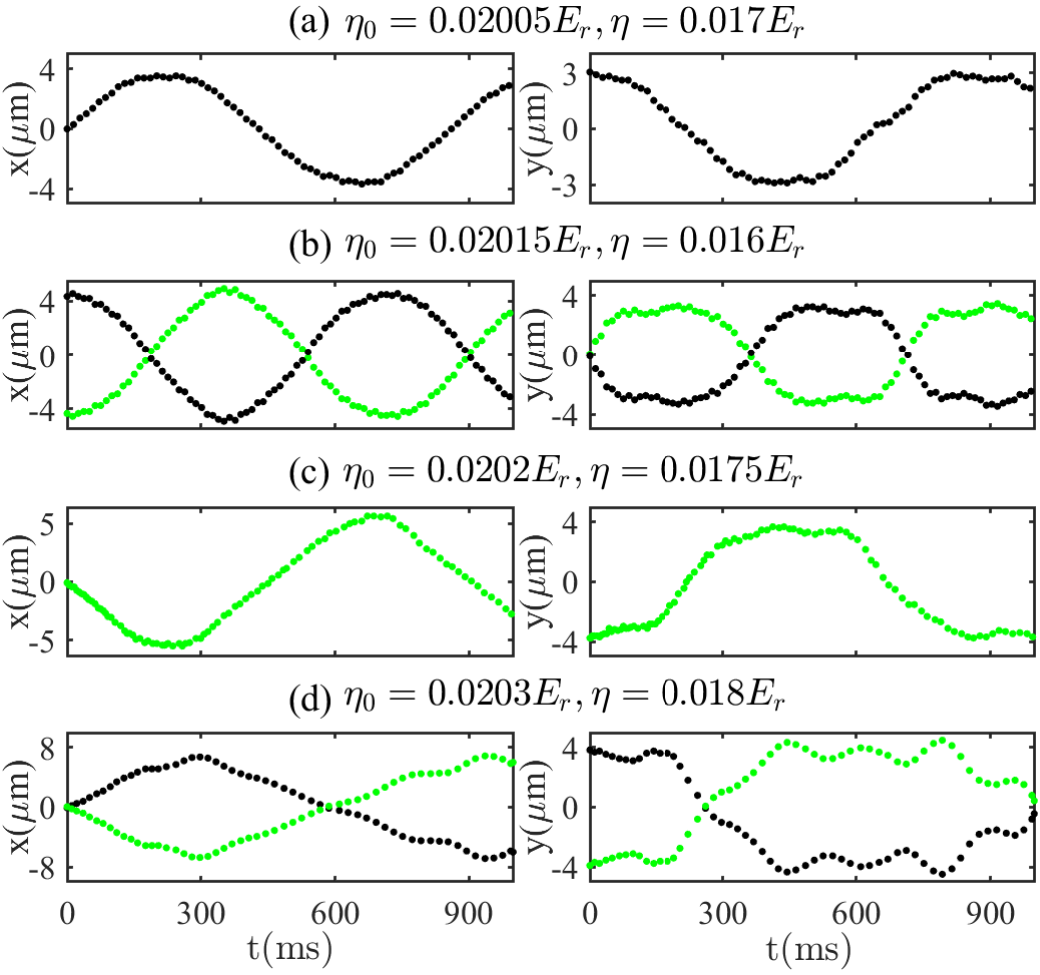}
\caption{Time evolution of vortex trajectories along the $x$ and $y$-directions. (a), (b), (c), and (d) correspond to the changes of vortices' positions marked by green or black circles in Fig.~\ref{fig:vortex3456}, respectively.}
\label{fig:motion}
\end{figure} 
 
In the synthetic magnetic field experiment of $^{87}$Rb BEC, the relation between the detuning gradient and the applied magnetic field gradient is $\nabla B=- \eta k_{r}/(\text g   \mu _{B})$, where $\text g$ is the Land\'{e} $\text g$-factor. Adjusting the magnetic field gradient will reflect variations of the maximum rotation angle $\theta_{max}$ and rotation period $T $, which can be measured by direct imaging in real time of the total density distribution of the two-component BEC. Thus, the system could be used as a gradient magnetometer by relating changes in rotation angle and period to variations of the modulus of the magnetic field gradient. In this experiment, two Raman lasers $\lambda = 804.1$nm are separated by a $\pi/2$ angle and the Zeeman shift for the $ F = 1$ ground state in the low-field regime is approximately $0.7$ MHz/Gauss \cite{steck2001rubidium}. Through parameter conversion, we estimate the magnetic field gradient as: $\nabla B=5.33\times 10^{-4} -8.99\times 10^{-5} \ln_{}{(500\theta_{max}+17)  }$Tesla/cm for $(0.00915<\eta \le 0.02)$; $\nabla B=2.06\times 10^{-5} \ln_{}{(T -136)  }$Tesla/cm for $(0\le \eta <0.00915)$. Additionally, for the angle measurement ($0 < \theta_{max} < \pi/2$), the estimated sensitivity of gradient magnetic field ranges $2.64×10^{-4}\sim 9.95×10^{-7}  $Tesla/cm per degree. Similarly, for the period measurement ($137 < T < 588$ms), the sensitivity ranges $2.06×10^{-5}\sim 4.58×10^{-8}  $Tesla/cm per period. The sensitivity increases significantly near the dynamical critical point $\eta_{c}$ of the twin vortices, achieving optimal performance.

Interestingly, vortex lattice states with different nucleations also exhibit stable periodic rotational dynamics when the quenched detuning gradient deviates from the initial value ($\eta<\eta_{0}$). Figures~\ref{fig:vortex3456} and \ref{fig:motion} demonstrate the dynamics of different vortex lattice states. For the triangular vortex lattices, three vortices are localized at $(0,3.04)\mu$m, $(3.04,-1.52)\mu $m, and $(-3.04,-1.52)\mu $m at the ground state ($\eta_{0} =0.02005E_{r}$), respectively. After quenching to $\eta = 0.017E_{r}$, vortices undergo clockwise rotation. At the quench time $\frac{1}{3}T  =288$ms, their positions become $(3.04, -1.52)\mu$m, $(-3.04, -1.52)\mu$m, and $(0, 3.04)\mu$m. After $ t=864$ms, the three vortices return to their initial positions. For the rhombic vortex lattices initially located at $(-4.34,0)\mu$m, $(0,3.62)\mu $m, $(4.34,0)\mu $m, and $(0,-3.62)\mu $m with $\eta_{0} =0.02015E_{r}$, they rotate to positions $(4.34,0)\mu$m, $(0,-3.62)\mu $m, $(-4.34,0)\mu $m, and $(0,3.62)\mu $m at $\frac{1}{2}T  =360$ms, after quenching to $\eta = 0.016E_{r}$. The pentagonal vortex lattices initially occupy $(0,-3.76)\mu$m, $(-4.92,-1.52)\mu $m, $(-2.67,3.54)\mu $m, $(2.67,3.54)\mu $m, and $(4.92,-1.52)\mu $m for $\eta_{0} = 0.0202E_{r}$. When quenching $\eta = 0.0175E_{r}$, vortices exhibit clockwise rotation with period $T =920$ms. Their positions shift to $(-4.92,-1.52)\mu $m, $(-2.67,3.54)\mu $m, $(2.67,3.54)\mu $m, $(4.92,-1.52)\mu $m, and $(0,-3.76)\mu$m at quench time $\frac{1}{5}T  =184$ms. When the initial detuning gradient $\eta_{0} = 0.0203E_{r}$, the hexagonal vortex lattices are localized at $(0,-3.83)\mu $m, $(-4.71,-3.11)\mu $m, $(-4.71,3.11)\mu $m, $(0,3.83)\mu $m, $(4.71,3.11)\mu $m, and $(4.71,-3.11)\mu$m at $t=0$ms. After quenching to $\eta = 0.018E_{r}$, their positions become $(-4.71,-3.11)\mu $m, $(-4.71,3.11)\mu $m, $(0,3.83)\mu $m, $(4.71,3.11)\mu $m, $(4.71,-3.11)\mu$m, and $(0,-3.83)\mu $m at $\frac{1}{6}T  =185$ms. The rotational dynamics of these topological structures could facilitate the encoding and transmission of quantum information and hold promise for potential applications in quantum communication.

\begin{figure}[t!]
\centering
\includegraphics[width=1\linewidth]{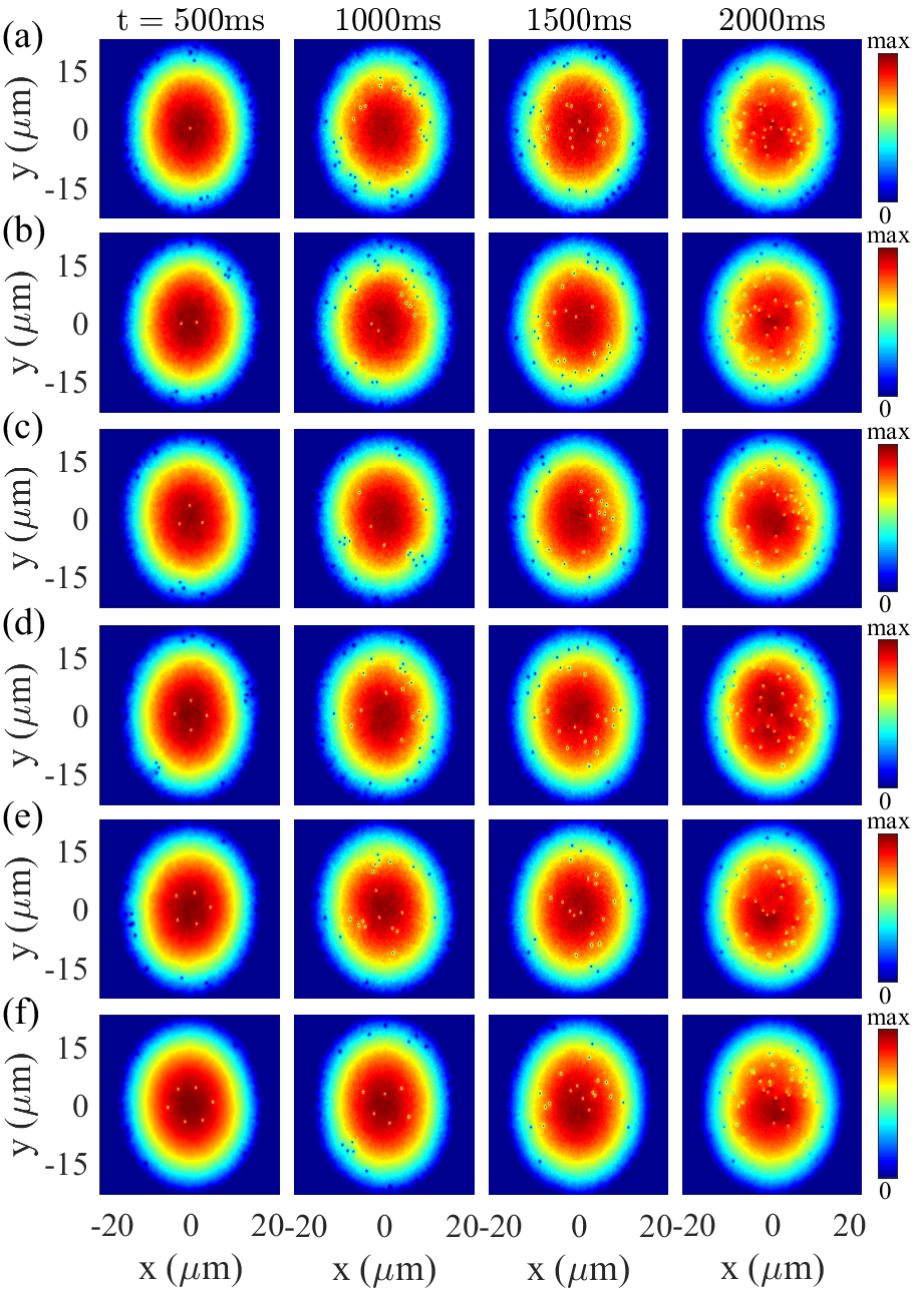}
\caption{Evolution of total density of single (a), two (b), triangular (c), rhombic (d), pentagonal (e), and hexagonal vortex lattices (f) at different quench times for quenching detuning gradients $\eta= 0.023E_{r}  $, respectively. }
\label{fig:vortices}
\end{figure}


The detuning gradient $\eta$ controls the emergence of vortices in synthetic magnetic field experiment. A straightforward method to generate more vortices is to increase this detuning gradient. Next, we consider the dynamics of different vortex configurations when the quenched detuning gradient is beyond its initial value. Figure~\ref{fig:vortices} displays the density distributions of vortex lattices at different quench times for $\eta = 0.023E_{r}$. Initially, the existing vortices also undergo rotational motion. With the increase of quench time, new vortices appear on the periphery of the condensate, exhibiting counterclockwise rotation, and then moving irregularly to the center of the condensate. Significant vortex proliferation occurs within $500$-$1000$ms for most vortex lattice states, while for hexagonal vortex lattices, this primarily happens between $1000$-$1500$ms. The vortex numbers of each lattice state approach dynamic stability within $1500$-$2000$ms.

\section{Conclusion}
\label{sec-conclusion}

In summary, we have systematically investigated the quenching dynamics of vortex states in SOC BECs under position-dependent detuning. Our numerical simulations reveal that quenching the detuning gradient below its initial value induces stable periodic rotation in both twin vortices and larger vortex lattices, with coherence sustained for up to 1000 ms. Notably, twin vortices exhibit two distinct dynamical regimes: scissors-like oscillation and unidirectional rotation, depending on the gradient strength. In contrast, quenching the gradient above this threshold triggers the nucleation of additional vortices.

Most significantly, by fitting the numerical results from the GP equations, we establish a quantitative relation between the vortices' maximum rotation angle (or period) and the applied detuning gradient. This relation, combined with the fact that the rotation angle and period can be directly read out via fluorescence imaging, establishes this system as a promising platform for quantum sensing. As a concrete example, for a gradient magnetometer operating near the dynamical critical point of the twin vortex states, we estimate a sensitivity on the order of  $10^{-8}$Tesla/cm.

Beyond sensing, our findings suggest intriguing potential for quantum information processing. The distinct states and controllable $\pi$-phase jumps in rotating twin vortices could be encoded as logical qubit states  (e.g., $\left | 01 \right \rangle $, $\left | 10 \right \rangle $), with their tunable dynamics offering a mechanism for implementing dynamically controllable quantum gates. Furthermore, the long-lived, periodic stability of the vortex lattices highlights their potential for quantum memory applications.

Our work provides fundamental insights into the non-equilibrium dynamics of vortices in synthetic magnetic fields and should motivate further experimental exploration. Future theoretical studies could extend this framework to investigate vortex dynamics under the influence of gravity \cite{PhysRevA.101.063616}, within optical lattices \cite{NP2025,WOS:000937133200011} or cavities \cite{PhysRevResearch.7.013170}, and explore optimization via artificial intelligence \cite{WOS:001434710500010}.

\begin{acknowledgments}
J.W., Z.F., and Y.L. are supported by the National Key Research and Development Program of China (Grant No. 2025YFF0515201), the Joint Funds of the National Natural Science Foundation of China (Grant No. U25D8014), the National Natural Science Foundation of China (Grant No. 11774093), the Natural Science Foundation of Shanghai (Grant No. 23ZR1418700), and the Innovation Program of Shanghai Municipal Education Commission (Grant No. 202101070008E00099).
\end{acknowledgments}


\bibliography{ref}

\end{document}